\documentclass[twocolumn,preprintnumbers,amsmath,amssymb]{revtex4}
\usepackage{graphicx}
\usepackage{dcolumn}
\usepackage{bm}

\begin{document}


\title{
Nonequilibrium chiral dynamics by the time dependent variational approach\\
with squeezed states}

\author{N. Ikezi$^{1,2}$}
\author{M. Asakawa$^2$
 \email{yuki@ruby.scphys.kyoto-u.ac.jp}}
\author{Y. Tsue$^3$}
\affiliation{
$^1$Department of Physics, Nagoya University, Nagoya 464-8602, Japan \\
$^2$Department of Physics, Kyoto University, Kyoto 606-8502, Japan \\
$^3$Physics Division, Faculty of Science, Kochi University, Kochi
780-8520, Japan}

\date{Received 10 October 2003}

\begin{abstract}
 We investigate the inhomogeneous chiral dynamics of the O(4) linear
 sigma model in 1+1 dimensions using the time dependent 
 variational approach in the space spanned by the squeezed states. 
 We compare two cases, with and without the Gaussian approximation
 for the Green's functions.
 We show that mode-mode correlation plays a decisive role in the
 out-of-equilibrium quantum dynamics of domain formation and
 squeezing of states.
\end{abstract}


\maketitle
The possibility of the formation of the disoriented chiral condensate
(DCC) in high energy heavy ion collisions has been extensively studied 
with various methods. In classical approximation
 \cite{ref:classical01,ref:classical02},
it has been shown that the amplification of long wavelength
modes of the pion fields takes place when the system starts
with the nonequilibrium initial condition, 
quench initial condition \cite{ref:classical01,ref:classical02}.
In addition to the amplification, spatial correlation of the fields
has been also shown to grow.

Although the classical approximation is expected to work well
in incorporating nonequilibrium aspects of the system
when pion density is large,
it is still desirable to include quantum effects.
In fact, investigations in this direction have been also carried out
extensively with the Hartree approximation, the large $N$ approximation, 
and so on \cite{boyanovski01,cooper01}.
In most of the previous studies which include quantum effects, however,
it has been assumed that the system is spatially homogeneous.
Problems such as insufficient thermalization
at late times and impossibility to describe domain structures
have been recognized. 
It has not been conclusive whether there is a chance for the
correlations to grow through nonequilibrium time evolution.

There are at least two ways for possible improvement.
One is to include higher order quantum corrections and
the other is to accommodate spatial inhomogeneity.
We will pursue the latter in this paper. 
Recently, the dynamics of spatially inhomogeneous system
has been studied by several groups quantum mechanically
\cite{berges,cooper02,smit,bettencourt} and
it has been shown that the thermalization of the quantum fields
can occur. In these works, the Gaussian approximation, 
in which the Green's functions are assumed to be diagonal
in momentum space, has been adopted because of computational reasons.
Physically, it corresponds to ignoring correlations 
between modes with different momenta, and
under the approximation different modes can interact only through the mean
fields. However, it is possible that the direct coupling of modes
through the off-diagonal correlations is important for
the time evolution of the system when the system does not possess
translational invariance. 
To see if such an effect is substantial, 
we study the dynamics of chiral phase transition in spatially
inhomogeneous systems with off-diagonal components of the Green's
function in momentum space fully taken into account.

In this paper, we take the O(4) linear sigma model as a low energy
effective theory of QCD and apply the method of the time dependent variational
approach (TDVA) with squeezed states.
This method was originally developed by Jackiw and Kerman as 
an approximation in the functional Schr\"odinger approach
\cite{JK} and later it was shown to be equivalent to TDVA with squeezed
states by Tsue and Fujiwara \cite{TF}.

In this approach,
the trial state is a squeezed state
\begin{eqnarray}
 | \Phi (t) \rangle & = & \prod_{a}| \Phi_{a} (t) \rangle, \nonumber \\
 | \Phi_{a} (t) \rangle & =  & \exp \bigl[S_a(t)\bigr]
 \cdot N_a (t) \cdot
 \exp \bigl[T_a (t)\bigr] |0 \rangle, \nonumber \\
 S_a (t) & = & i \int d \vec{x} 
 \bigl[ C_{a}(\vec{x},t) {\phi_a}(\vec{x}) 
 - D_{a}(\vec{x},t){\pi_a}(\vec{x}) \bigr], \nonumber \\
 T_a (t) & = & \int d \vec{x} d \vec{y} 
 \phi_{a}(\vec{x}) \Bigl[
 - \frac{1}{4} \bigl[ G_{a} ^{-1}(\vec{x},\vec{y},t) \nonumber \\ 
 & & - G_{a}^{(0)-1}(\vec{x},\vec{y}) \bigr]
 + i \Pi_{a}(\vec{x},\vec{y},t) \Bigr] \phi_{a} (\vec{y}) 
 \label{squeezedstate},
\end{eqnarray}
where $a$ runs from 0 to 3. $a=0$ is for the sigma field and
$a = 1-3$ are for the pion fields. $|0 \rangle$ is the reference vacuum
and 
$G_{a}^{(0)}(\vec{x},\vec{y}) = 
\langle 0|\phi_a(\vec{x})\phi_a(\vec{y})|0 \rangle $. 
$\phi_a(\vec{x})$ and $\pi_a(\vec{x})$ are the field operator and
conjugate field operator for the field $a$, respectively.
$C_{a}(\vec{x},t)$, $D_{a}(\vec{x},t)$,
$G_{a}(\vec{x},\vec{y},t)$, and $\Pi_{a}(\vec{x},\vec{y},t)$ 
are the mean field variables for $\phi_{a}$ field at
$\vec{x}$ and $t$, the canonical conjugate variable for the mean field,
the quantum correlation for $\vec{x} \neq \vec{y}$ 
(fluctuation around the mean field for $\vec{x} = \vec{y}$), and
the canonical conjugate variable for $G_{a}(\vec{x},\vec{y},t)$, respectively, 
and all of them are real functions.
$N_a (t)$ is a normalization constant. $T(t)$ is an operator
that describes the squeezing, and if $T_a (t)$ is set to 0, 
$| \Phi_{a} (t) \rangle $ is reduced to a coherent state with the
expectation values of $\phi(\vec{x},t)$ and $\pi(\vec{x},t)$ given by
$C_{a}(\vec{x},t)$ and $D_{a}(\vec{x},t)$, respectively.

The Hamiltonian $H$ of the O(4) linear sigma model is given by
\begin{eqnarray}
 H & = &\int d \vec{x} \sum_{a=0}^3 \biggl \{ \frac{1}{2}\pi_a(\vec{x})^2 +
 \frac{1}{2}\vec{\nabla}\phi_a(\vec{x})\cdot
 \vec{\nabla}\phi_a(\vec{x}) \nonumber \\
 & & + \lambda \bigl[ \phi_a(\vec{x})^2 - v^2 \bigr]^2
 - h\phi_0(\vec{x}) \biggr \}.
 \label{hamiltonian:o4}
\end{eqnarray}
As shown in Eq. (\ref{squeezedstate}),
the trial state is specified by 
$C_{a}(\vec{x},t)$, $D_{a}(\vec{x},t)$, 
$G_{a}(\vec{x},\vec{y},t)$, and $\Pi_{a}(\vec{x},\vec{y},t)$.
Their time evolution is determined through the time dependent variational
principle: 
\begin{equation}
 \delta \int dt \langle \Phi (t) | i \frac{\partial}{\partial t}
  - H | \Phi (t) \rangle = 0. \label{TDVP}
\end{equation}
In this approach, correlation between different modes in momentum space
arises through the scattering of quanta caused by 
the nonlinear coupling term in the model Hamiltonian
even if there is initially no such correlation.
This can be seen best from the following equations of motion in momentum space,
\begin{eqnarray}
 \ddot{C}_{a}(\vec{k},t) & = & - \vec{k}^{2} - \mathcal{M}_{a}^{(1)}(\vec{k},t)
 ,\nonumber \\
\dot{G}_{a}(\vec{k},\vec{k}',t) & = & 2 \langle \vec{k} | \left[
 G_{a}(t) \Pi_{a}(t) + \Pi_{a}(t) G_{a}(t) \right] | \vec{k}' \rangle ,
 \nonumber \\ 
\dot{\Pi}_{a}(\vec{k},\vec{k}',t)& = & \mbox{$\frac{1}{8}$} \langle \vec{k} |
 G_{a}^{-1}(t) G_{a}^{-1}(t) | \vec{k}' \rangle
 - 2 \langle \vec{k} | \Pi_{a}(t) \Pi_{a}(t) | \vec{k}' \rangle \nonumber \\
 & &- \vec{k}^{2} \delta ^{3} ( \vec{k} - \vec{k}' )
 - \mbox{$\frac{1}{2}$} \mathcal{M}_{a}^{(2)} (\vec{k}-\vec{k}',t), \nonumber \\
\mathcal{M}_{a}^{(1)}(\vec{k},t) & = &
 \Bigl[- m^{2} + 4 \lambda C_{a}^{2}(\vec{k},t)
 + 12 \lambda G_{a}(\vec{k},\vec{k},t) \nonumber \\
 & & + 4 \lambda \sum_{b (\neq a)} \bigl( C_{b}^{2}(\vec{k},t)
 + G_{b}(\vec{k},\vec{k},t) \bigr) \Bigr] C_{a}(\vec{k},t) \nonumber \\
 & &- h\delta_{a0} , \nonumber \\
\mathcal{M}_{a}^{(2)}(\vec{k},t) & = &
 - m^{2} + 12 \lambda C_{a}^{2}(\vec{k},t)
 + 12 \lambda G_{a}(\vec{k},\vec{k},t) \nonumber  \\
 & & + 4 \lambda \sum_{b (\neq a)} \left( C_{b}^{2}(\vec{k},t)
 + G_{b}(\vec{k},\vec{k},t) \right),  
\label{eom}
\end{eqnarray}
where $m^2  =  4 \lambda v^2$, and $C_a(\vec{k},t)$, 
$G_a(\vec{k},\vec{k}',t)$,
and $\Pi_a(\vec{k},\vec{k}',t)$  are
the mean fields for the $\phi_a$ field with momentum $\vec{k}$,
the correlation between modes with momenta $\vec{k}$ and $\vec{k}'$
for $\vec{k} \neq \vec{k}'$
(the quantum fluctuation around the mean field for $\vec{k} = \vec{k}'$),
the canonical conjugate variable for $G_a(\vec{k},\vec{k}',t)$,
respectively. In Eq. (\ref{eom}), we have used the notation,
\begin{equation}
 \langle \vec{k} | H(t) I(t) | \vec{k}' \rangle =
  \int \frac{d \vec{k}''}{(2\pi)^{3}}
  H(\vec{k},\vec{k}'',t) I(\vec{k}'',\vec{k}',t).
\end{equation}
In the Gaussian approximation,
$G_{a}(\vec{k},\vec{k}',t)$ and $\Pi_{a}(\vec{k},\vec{k}',t)$
are set to zero for $\vec{k}\neq \vec{k}'$ and
correlations between different
modes in momentum space are ignored.
However, $\mathcal{M}^{(2)}(\vec{k}-\vec{k}',t)$ in Eq. (\ref{eom}),
which originates from the four-point interaction terms in the O(4)
linear sigma model, couples modes with different momenta 
and correlations between them develop even if initially 
there exists no correlation among them.

In numerical calculation, we have assumed the one-dimensional
spatial dependence for the mean fields and the Green's functions
for computational simplicity.
In addition, we have imposed the
periodic boundary condition for the mean fields and the Green's
functions. 
We have carried out calculation on a lattice
with the lattice spacing $d=1.0$ fm and the total length
$L=64$ fm,
which leads to the momentum cutoff $\Lambda = 1071$ MeV.
The parameters $\lambda$, $v$, and $h_0$ are determined so that
they give the pion mass $M_{\pi} = 138$ MeV,
the sigma meson mass $M_{\sigma}=500 $ MeV,
and the pion decay constant $f_{\pi} = 93 $ MeV in the vacuum
following the prescription given in Ref. \cite{TKI},
and we have obtained $\lambda = 3.44$, $v = 110$ MeV,
and $h_0 = (103 ~\rm{MeV})^{3}$ .

There are several scenarios for the DCC formation in high energy heavy
ion collisions. Here we adopt the quench scenario. 
In this scenario, the chiral order parameters remain 
around the top of the Mexican hat potential
after the rapid change of the effective potential from the chirally
symmetric phase to the chirally broken phase.
In order to take this situation into account, we have used the following
initial condition; at each lattice point, the mean field variable
for the chiral fields and their conjugate variables
$C_a(\vec{x},0)$ and $D_a(\vec{x},0)$ are randomly distributed
according to the Gaussian form with the following parameters \cite{AM},
\begin{eqnarray}
 \langle C_a(\vec{x},0) \rangle &=& 0, \nonumber \\
 \langle C_a(\vec{x}, 0)^2 \rangle
 - \langle C_a(\vec{x}, 0) \rangle ^2 &=& \delta^2 , \nonumber \\
 \langle D_a(\vec{x}, 0) \rangle &=& 0 , \nonumber \\
 \langle D_a(\vec{x}, 0)^2 \rangle
 - \langle D_a(\vec{x},0) \rangle ^2 &=& \frac{D}{d^2} \delta ^2 ,
 \label{initial:mf4}
\end{eqnarray}
where $D = 1$ is the spatial dimension and
$\delta$ is the Gaussian width.
We shall use $\delta = 0.19 v$ in the following calculations.
In relating the Gaussian widths of $C_a(\vec{x}, 0)$ and $D_a(\vec{x}, 0)$,
we have taken advantage of the virial theorem \cite{AM}.

As for the initial conditions for the quantum fluctuation and correlation,
we have assumed that their values are those realized in the case
where each state in momentum space were in a
coherent state with a degenerate mass $m_{0}$ for the sigma meson and
pions, namely, 
\begin{eqnarray}
 & & G_a(\vec{x}, \vec{y},0) =
 \int_{0}^{\Lambda} \frac{d \vec{k}}{(2 \pi)^{3}} \frac{1}{2 \omega_k}
 e ^{i \vec{k}\cdot(\vec{x}- \vec{y})} , \nonumber  \\
 & & \Pi_a(\vec{x}, \vec{y}, 0) = 0 , \label{coh_fluc}
\end{eqnarray}
where $\omega _{k} = \sqrt{m_{0}^2 + \vec{k}^2}$.
We adopt $m_{0} = 200$ MeV. 
As shown above, the Green's functions $G_{a}(\vec{x},\vec{y},0)$
and $\Pi_a(\vec{x},\vec{y},0)$ are initially diagonal in momentum space. 
The off-diagonal components appear in the course of the time evolution
of the system due to the direct mode-mode coupling induced by
$\mathcal{M}^{(2)}$ in Eq. (\ref{eom}).

We have carried out two sets of numerical calculations.
In one case, we have taken into account
all components of the two-point Green's functions
(case I), while in the other case only the diagonal 
components of the two-point Green's functions (case II) were
included in the calculation as in most of the preceding works.

\begin{figure}[tbp]
               \resizebox{72.5mm}{!}{\includegraphics{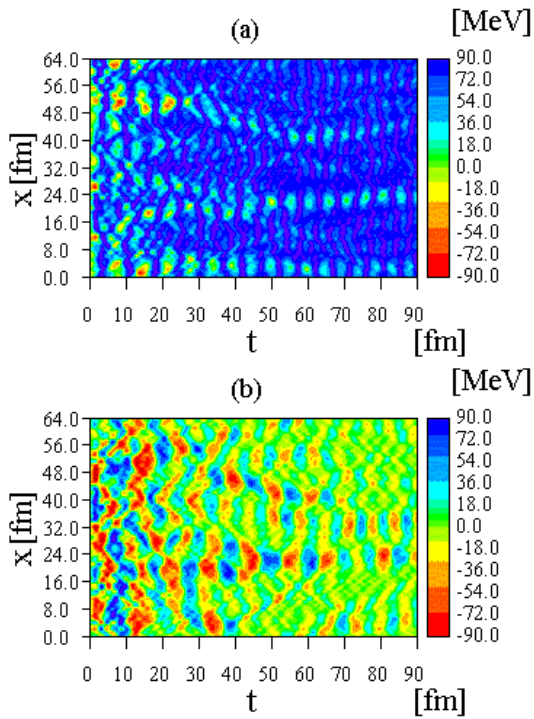}}
                \caption{ (Color online) Time evolution of the
	 mean fields in case I. (a) and (b) are for the
	 sigma field and the third component of the pion field,
         respectively.}
         \label{fig:meanfields01}
\end{figure}

In Fig.~\ref{fig:meanfields01}, we show the time evolution of the mean
fields for the sigma and the third component of the pion fields
($\phi_0$ and $\phi_3$, respectively) obtained
with the calculation including all components of the
Green's functions (case I). 
As a whole, the expectation value of the sigma field approaches a constant.
On the other hand, that of the pion field oscillates around zero
and shows a domain structure with long correlation length.
This is the formation of DCC domains.
It is observed that the domain structure continues to grow till as late as
$t \sim 40$ fm.

\begin{figure}[tbp]
               \resizebox{72.5mm}{!}{\includegraphics{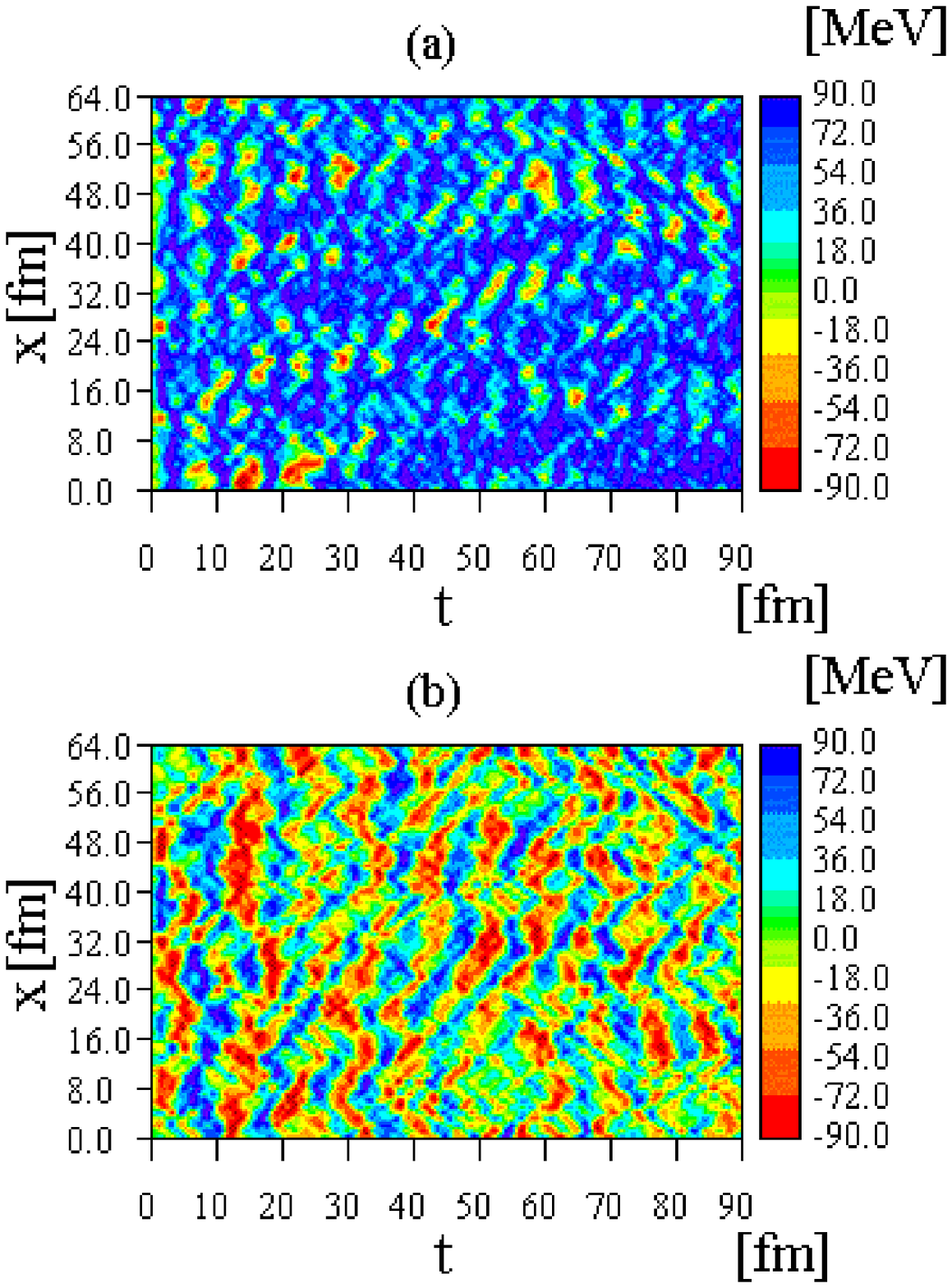}}
                \caption{(Color online) Time evolution of mean
	 fields in case  II. (a) and (b) are for the
	 sigma field and the third component of the pion field,
         respectively.}
         \label{fig:meanfields02}
\end{figure}

In Fig.~\ref{fig:meanfields02}, we show the time evolution of the same
mean fields obtained with the Gaussian approximation, i.e.,
without the off-diagonal components of the Green's functions (case II). 
At the beginning, 
the 
behavior of the mean fields is similar to that in case I.
However, after a few femtometers,
there appears a clear difference between the
two cases. In case II, short range fluctuation is dominant and
no long length correlation is observed. No qualitative
change in the behavior of the mean fields takes place in case II
after a few femtometers.
This tells us that the mode-mode correlation plays a decisive role in
the formation of DCC domains.

To examine the growth of spatial correlation of the pion fields
more quantitatively, 
we define the following spatial correlation function $C(r, t)$,
\begin{equation}
 C(r, t) = \frac{\int 
           {\vec{\phi} }(\vec{x}) \!\cdot\! {\vec{\phi}}(\vec{y})
           \delta (|\vec{x} - \vec{y}| - r) d\vec{x} d\vec{y} }
           {\int | {\vec\phi}(\vec{x}) | | {\vec\phi}(\vec{y}) |
           \delta (|\vec{x} - \vec{y}| - r) d\vec{x} d\vec{y} },
\end{equation}
where ${\vec{\phi} }(\vec{x})\!\cdot\! {\vec{\phi}}(\vec{y})
= \sum_{i=1}^{3} {\phi_i}(\vec{x}) {\phi_i}(\vec{y})$
and $|{\vec\phi}(\vec{x}) | =
\sqrt{ \sum_{i=1}^{3} {\phi_i^2}(\vec{x}) }$.

\begin{figure}[t]
\centerline{\includegraphics[width=65mm,]{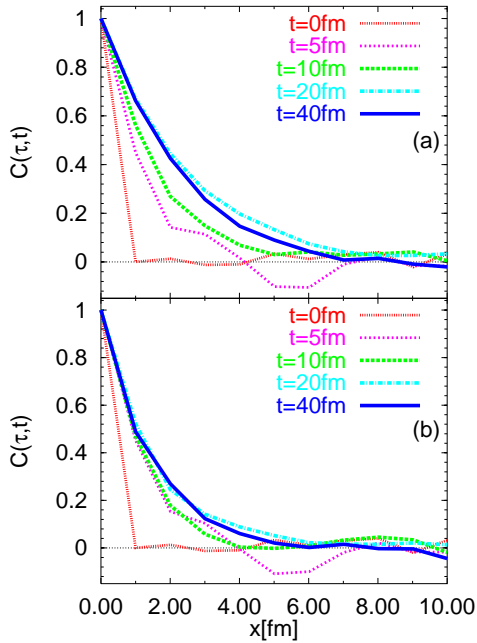}} 
               \caption{(Color online)
                 The spatial correlations for the pion fields
                 in case I (a) and case II (b).}
               \label{fig:correlation1}
\end{figure}

In Figs.~\ref{fig:correlation1}(a) and \ref{fig:correlation1}(b),
 we show
this spatial correlation function in cases I and II, respectively.
The correlation functions are calculated by taking average over 10 events.
Substantial generation of the correlation takes place
after the typical time scale of the initial rolling down of the chiral
fields, say, a few femtometers
in case I, while the growth of the spatial
correlation ends in case II by $t\sim5$ fm.
The domain formation of DCC shown in Fig.~\ref{fig:meanfields01}(b) 
and this growth of spatial correlation shown in
Fig.~\ref{fig:correlation1}(a) beyond the rolling down time scale may be
related to the parametric resonance 
\cite{mrowczynski95,hiro-oka00}. 
We are currently investigating this possibility.
\begin{figure}[tbp]
      \resizebox{85mm}{!}{\includegraphics{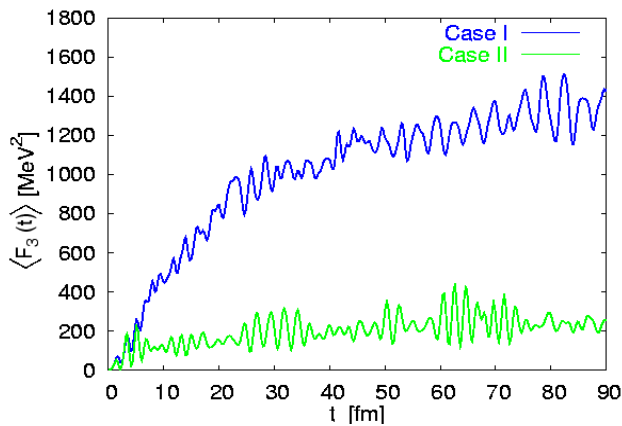}}
                \caption{(Color online)
 The time evolution of the spatially averaged
 quantum fluctuation of the third component of the pion field,
 $\langle F_3(t) \rangle_{\rm space}$ for  cases I and II. }
    \label{fig:quantum_fluctuation}
\end{figure}

Next we show the time evolution of the quantum fluctuation, which is
represented by the same-point Green's function $G_{a}(\vec{x}, \vec{x}, t)$.
We define the spatially averaged fluctuation
function $\langle F_a(t)\rangle _{{\rm space}}$ at time $t$ as
\begin{equation}
  \langle F_{a}(t) \rangle_{{\rm space}} = \frac{1}{V}
  \int G_{a}(\vec{x} , \vec{x}, t) d{\vec{x}} ,
\end{equation}
where $V$ is the volume of the system.

In Fig.~\ref{fig:quantum_fluctuation}, 
we compare the time evolution of the spatially averaged quantum
fluctuation of the the third component of the pion field,
$\langle F_3 (t) \rangle_{{\rm space}}$ in the two cases.
We observe remarkable increase of quantum fluctuation
in case I, while only small amplification is seen in case II.
This increase also lasts until about $t = 40$ fm.
The comparison between case I and case II tells us
that the off-diagonal correlations take an important role
also for the enhancement of the quantum fluctuation.
We have in fact found the growth of off-diagonal components of the
Green's functions in momentum space.
Note that this phenomenon cannot be described unless
the off-diagonal correlation is introduced.
This has an important meaning also for the identical particle
correlation in high energy heavy ion collisions.
Usually it is assumed that identical particles with different
momenta are emitted independently. However, if there is quantum
correlation between two different modes, this assumption becomes
invalid, and it will be necessary to reformulate the theory for
the identical particle correlation.

In summary, we have studied the inhomogeneous chiral dynamics of the
O(4) linear sigma model
in 1+1 dimensions using TDVA with squeezed states.
We have compared two cases. 
One is a general case in which both the mean fields and the Green's functions
are inhomogeneous, and
the other is a case with the Gaussian approximation,
where translational invariance is imposed on the Green's functions.
We have shown for the first time
that the large correlated domains can be realized 
in the quench scenario with quantum mechanical treatment.
More specifically, we have shown that the large amplification of quantum
fluctuation and large domain structure emerge when all components
of the Green's functions are retained,
while only small quantum fluctuation and small and noisy
domain structure are seen in the case with the Gaussian approximation.
The dynamics in $1+2$ and $1+3$ dimensional cases is of great
interest and indispensable for the understanding of the DCC
formation in ultrarelativistic heavy ion collisions. We expect
more enhanced domain growth in $1+2$ and $1+3$ dimensional cases
because of the existence of more mode-mode correlations in such cases.
We plan to confirm it by actual numerical calculation.

N.I. thanks the nuclear theory group at Kyoto University
for encouragement. M.A. is partially
supported by the Grants-in-Aid of the Japanese Ministry of Education,
Science and Culture, Grants Nos.~14540255. 
Y.T. is partially
supported by the Grants-in-Aid of the Japanese Ministry of Education,
Science and Culture, Grants No.~13740159 and 15740156.
Numerical calculation was carried out at
Yukawa Institute for Theoretical Physics at Kyoto University and
Tohoku-Gakuin University. We thank T. Otofuji for making it possible
to use the computing facility at Tohoku-Gakuin University.

\end{document}